\documentclass[fleqn,11pt]{olplainarticle}
\usepackage{siunitx}
\usepackage[english]{babel}
\usepackage{hyperref}
\usepackage{soul}
\usepackage{comment}
\usepackage{lineno}
\usepackage{colortbl}
\usepackage{booktabs}
\usepackage{rotating}
\usepackage{enumitem}
\usepackage{url}
\usepackage{float}
\usepackage{makecell}
\usepackage{fdsymbol}
\hypersetup{colorlinks=true, linkcolor=black, citecolor=black, urlcolor=blue}
\setlist{noitemsep,topsep=-0.75em,parsep=0pt,partopsep=0pt}
\addto\extrasenglish{

}

\title{Balancing training load, rest and musculoskeletal injury risk: a mathematical modelling study in Thoroughbred racehorses}

\author[1,2$\clubsuit$]{Md Nurul Anwar}
\author[1,2$\clubsuit$]{Michael Pan}
\author[1]{Ashleigh V. Morrice-West}
\author[3]{Fatemeh Malekipour}
\author[4]{Peter Pivonka}
\author[2]{Jennifer A. Flegg}
\author[1]{R Chris Whitton}
\author[1]{Peta L. Hitchens}
\affil[1]{Equine Centre, Melbourne Veterinary School, Faculty of Science, The University of Melbourne, Werribee, VIC 3030, Australia}
\affil[2]{School of Mathematics and Statistics, Faculty of Science, The University of Melbourne, Parkville 3010, Victoria, Australia}
\affil[3]{Department of Biomedical Engineering, Faculty of Engineering and Information Technology, The University of Melbourne, Parkville, VIC 3010, Australia}
\affil[4]{School of Mechanical, Medical and Process Engineering, Queensland University of Technology, Brisbane, QLD 4000, Australia}
\affil[$\clubsuit$] {These authors contributed equally to this work}


\begin{abstract}
Musculoskeletal injuries (MSI) in Thoroughbred racehorses are a leading cause of death and premature retirement in racehorses and are heavily influenced by training practices. Greater distances of high-speed galloping accumulated during racing campaigns are associated with MSI. Bone injury is the most common MSI, and understanding how training practices influence bone damage accumulation is critical for improving both horse welfare and racing outcomes. This study builds on an existing mathematical model of bone adaptation and damage to investigate the impact of different training programs on bone injury risk. Several training programs (three progressive, four race-fit, six rest programs and two with rest replaced by low-intensity training) were constructed to reflect representative practices undertaken by professional trainers in Victoria, Australia. Training programs varied in training volume, rest frequency and program duration. Lower volume training programs that included high-speed training, achieved sufficient bone adaptation with less accumulation of bone damage, and subsequently lower risk of bone failure. In addition, incorporating more frequent rests ($\ge$ 2 per year) and/or longer rest periods ($\ge$ 6 weeks) reduced bone damage due to the extended opportunity to remove and repair bone damage. These results provide an in-silico mathematical model of the bone’s response to training, demonstrating the effects of training programs on bone adaptation, damage formation and repair. The findings can guide the design of training programs that balance both bone adaptation and bone health throughout a horse’s racing career.
\end{abstract}

\begin{document}

\flushbottom
\maketitle

\section{Introduction}
Musculoskeletal injuries (MSI) are a major welfare concern in Thoroughbred racehorses. After decades of research identifying risk factors \citep{Hitchens2019Mar} and policies implemented to reduce injuries and fatalities, they remain an important concern to the industry. Fatal musculoskeletal injuries (FMIs) are the leading cause of race-related horse deaths \citep{Stover2017Feb,Boden2007Sep,morrice2025linkage}. Previous epidemiological studies have found associations between training practices and MSI. In particular, the speed of training and distances trained at high speeds have been linked to higher rates of MSI and fractures \citep{Verheyen2006Dec,Wong2023Association}. Alternatively, a matched case-control study showed that exercising horses $>44\ km$ at canter ($\le14\ m/s$) in combination with $6\ km$ galloping ($\ge14\ m/s$) per month was associated with increased risk of fracture \citep{Verheyen2006Dec}. \citet{MorriceWest2020Mar} surveyed trainers in Victoria, Australia, for their training practices, and clustered programs into three groups for progressive training (see \autoref{fig:training_program} for the structure of a typical training program) (fast and light, moderate, high volume) and four groups for race-fit training (low volume, medium volume, medium volume with increased high-speed training, and high volume). A subsequent study linked lower-volume progressive training programs to lower rates of MSI in two-year-old racehorses, but not mature racehorses \citep{Wong2023Association}. In mature racehorses, high-volume progressive training programs were associated with lower rates of MSI, but high-volume race-fit programs during a racing preparation were associated with a higher risk of FMI \citep{Wong2023Association}.

In addition to the volume of training, rest practices also influence risk of MSI. Postmortem studies identified higher micro-fracture densities in the third metacarpal subchondral bone of horses in training compared to resting horses \citep{Holmes2014Dec}. Lower bone resorption rates in racehorses in intensive training have also been observed, indicating inhibition of bone repair processes, and suggesting that time in rest is required for maintaining bone health \citep{Holmes2014Dec,whitton2018subchondral}. Optimal rest practices may be age dependent, as surveyed trainers who prefer shorter, more frequent rest periods had lower rates of injury in younger racehorses, but those preferring longer and more frequent rest periods in mature racehorses also had lower MSI risk in mature racehorses \citep{Wong2023Association}. This likely reflects the greater microdamage burden in mature horses. 

In contrast to statistical approaches, mechanistic models that incorporate processes related to bone remodelling and damage enforce physical and biological constrains presents in these systems. Modelling also provides an opportunity to reveal causal relationships between aspects of training programs and risk of bone failure. \citet{Hitchens2018Jun} developed the initial model of bone adaptation to investigate the response of bone volume fraction within the lateral condyle of the third metacarpal bone of Thoroughbred racehorses, to training and rest. This model was later expanded to include bone damage formation and repair processes, finding that joint stress and cycles (strides) per day were the key determinants of bone damage \citep{Pan2025Apr}. Mathematical models of bone fatigue life in racehorses have been applied to stride-level data in racehorses, revealing links between race distances, track surfaces and damage accumulation \citep{MorriceWest2022Jul,morrice2025impact}. However, mathematical models have yet to be used to understand how different aspects of a training program -- for example, the distances covered in training, rest frequency and rest duration -- are related to MSI risk.

This study aims to build on our existing mathematical model \citep{Pan2025Apr} to investigate the role of different training programs on bone injury risk in racehorses. To achieve this, we run simulations of our mathematical model under various progressive programs as horses enter and re-enter training following a rest period. These increase the horse's fitness prior to commencing racing. Maintenance training programs are the utilised once the horse is race-fit. We hypothesise that: (i) lower volume training programs with some high-speed training would lead to lower risk of bone failure; and (ii) more frequent, longer rests are linked to lower net bone damage. By simulating the effects of different training programs, this work provides a demonstration for the design of training programs that reduce bone damage while also adequately preparing bone for the loads induced by racing.

\section{Methods}
\begin{figure}
    \centering
    \includegraphics[width=\linewidth]{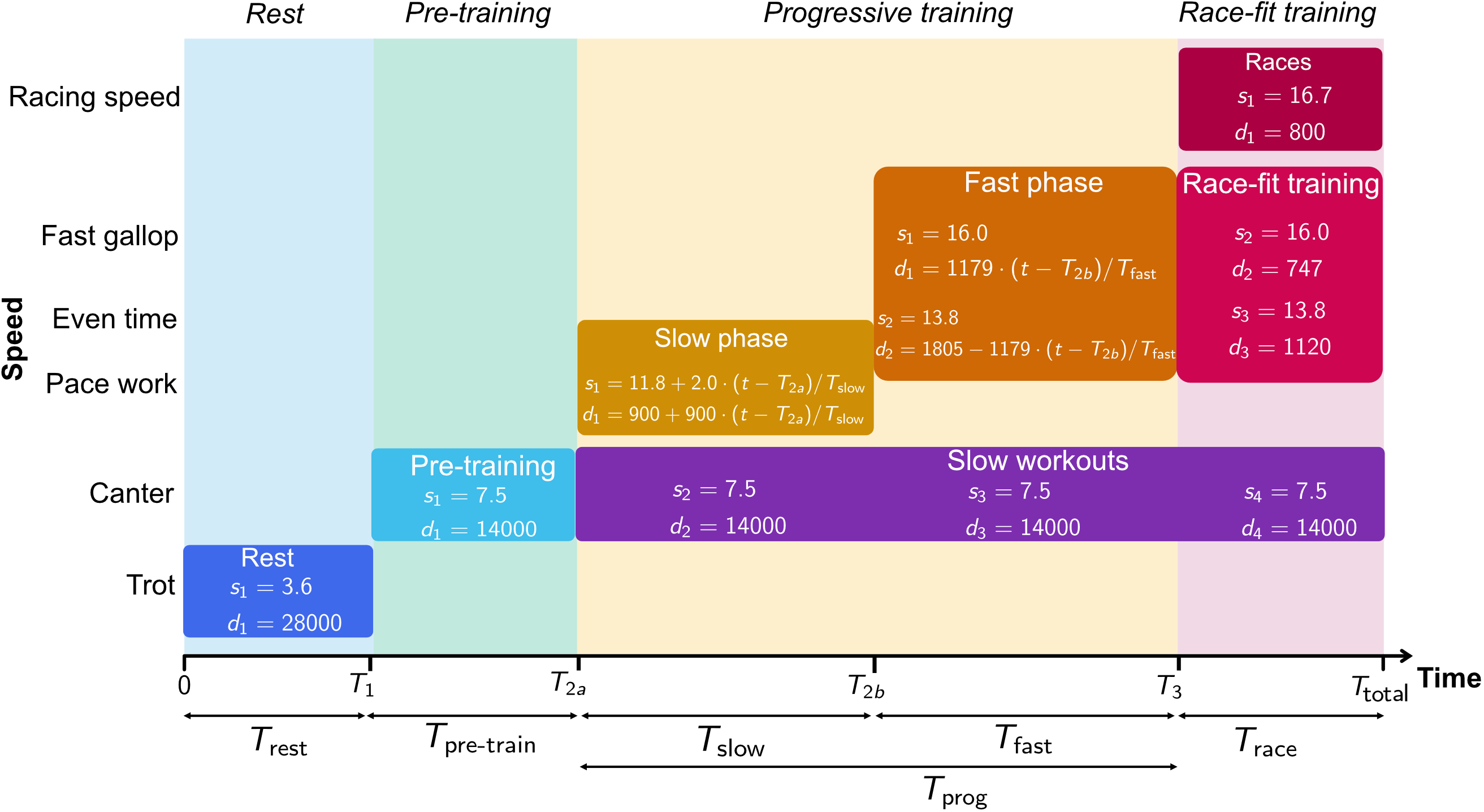}
    \caption{Schematic of a typical training preparation in Victoria, Australia. Speeds ($s_i$) are expressed in metres/second and corresponding distances ($d_i$) are expressed in metres/week. During each phase (coloured column), the indices $i$ are ordered from the fastest to slowest speed. For a typical training preparation, the durations of each training phase (in days) are $T_\text{rest} = 44$, $T_\text{pre-train} = 28$, $T_\text{slow} = 28$, $T_\text{fast} = 38$, $T_\text{prog} = T_\text{slow} + T_\text{fast} = 66$ and $T_\text{race} = 56$. The duration of the entire training program is $T_\text{total} = 194\ \si{days}$ with a rest frequency of 1.9 per year.}
    \label{fig:training_program}
\end{figure}

To investigate the impact of different training programs, we utilise a previously developed mathematical model developed by \citet{Pan2025Apr}. The full system of equations and parameters is given in \autoref{sec:mathematical_model}. Briefly, the model describes the coupled processes of bone adaptation and damage accumulation within a representative volume of interest (VOI) of the lateral condyle of the third metacarpal bone of a Thoroughbred racehorse, in response to training. The biological component of the model uses the joint stress $\sigma$ [MPa], strain rate $\dot{\varepsilon}$ [$\si{s^{-1}}$] and loading cycles per day $v_n$ [$\si{day^{-1}}$] as inputs. Within the model, these inputs can be connected to training intensity and volume by converting the speeds and distances covered per day into the input variables $\sigma$, $\dot{\varepsilon}$ and $v_n$. In general, a training preparation will include distances covered at both high and low speeds, and the speeds and volume of training evolve over time, with both speeds and distances increasing as the horse reaches race-level fitness. A training preparation is divided into periods of rest, pre-training, progressive training (split into a slow phase and a fast phase) and race-fit training (\autoref{fig:training_program}). \citet{Pan2025Apr} simulate the bone volume fraction and damage accumulation in response to a typical racehorse training program in Victoria, Australia. We use this as the default program from which other training programs are defined. The speeds and distances (as functions of time) used to simulate this training preparation are given in \autoref{fig:training_program}.

\subsection{Progressive training programs}

The external inputs to our model are adapted to match the clusters of progressive training programs (fast and light, moderate, and high-volume) derived in \citet{MorriceWest2020Mar}  based on surveyed work programs used by $n=66$ registered trainers in Victoria, Australia. We assume that the distance per week at the start of the slow phase (at speed 11.8 m/s; “three-quarter pace”) is equal to half the distance per week at speeds $\ge 13.8$ m/s (average of $13.3-14.3$ m/s) covered during the fast phase. This is linearly increased to the distance per week during the fast phase to approximate a continuous training program (see \citet{Pan2025Apr} for a full breakdown of a typical training preparation). Within the fast phase, horses begin galloping at 13.8 m/s, with the distance gradually replaced with the
higher speed of 16.0 m/s. A summary of the speeds and distances covered during each progressive training program cluster is given in \autoref{tab:progressive_properties}. All other aspects of the training program (i.e., rest, pre-training, and race-fit phases), including the speeds and distances for the slow workouts, are as defined in \autoref{fig:training_program}.

\begin{table}[]
    \centering
    \small
\caption{Durations and distances covered at various speeds for different progressive training programs undertaken by Thoroughbred racehorses \citep{MorriceWest2020Mar}. Other phases of training are simulated as described in \autoref{fig:training_program}. $s_i$ represents speeds and $d_i$ represents distances. Within each phase, the indices $i$ are ordered from the fastest to slowest speed (see Appendix \ref{speeds}). $T_{2a}=$ time passed until the start of slow progressive phase; $T_{2b}=$ time passed until the start of fast progressive phase.}
\label{tab:progressive_properties}
    \begin{tabular}{lccc}
    \toprule
 & \multicolumn{3}{c}{\textbf{Progressive program cluster}}\\ \midrule
 & Fast and light&Moderate &High-volume\\ \midrule
 \textbf{Program duration}& & &\\
         Total duration; $T_\text{prog}$ [days]&  63& 69&73\\
         Slow phase; $T_\text{slow}$ [days]& 
     30& 26&25\\
 Fast phase; $T_\text{fast}$ [days]& 33& 43&48\\ \midrule
 \textbf{Slow phase}& & &\\
 Total distance galloped [m]& 3614& 5170&7305\\
 Speed; $s_1$ [m/s]& $11.8 + 2.0 \big(\frac{t-T_{2a}}{T_\text{slow}}\big)$& $11.8 + 2.0\big( \frac{t-T_{2a}}{T_\text{slow}}\big)$&$11.8 + 2.0 \big(\frac{t-T_{2a}}{T_\text{slow}}\big)$\\[4pt]
 Distance covered per week; $d_1$ [m/wk]& $562 + 562 \big(\frac{t-T_{2a}}{T_\text{slow}}\big)$& $928 + 928\big(\frac{t-T_{2a}}{T_\text{slow}}\big)$&$1364 + 1364\big( \frac{t-T_{2a}}{T_\text{slow}}\big)$\\ \midrule
 \textbf{Fast phase}& & &\\
 Total distance covered [m]& 5300& 11400&18700\\
 \quad at 16.0 m/s& 2300& 3800&3700\\
 \quad at 13.8 m/s$^\text{a}$& 3000& 7600&
 15000\\
 Distance covered per week [m/wk] ($d_1+d_2$)& 1124& 1856&2727\\
 \quad at 16.0 m/s; $d_1$& $976 \big(\frac{t-T_{2b}}{T_\text{fast}}\big)$& $1237 \big( \frac{t-T_{2b}}{T_\text{fast}}\big)$&$1079\big( \frac{t-T_{2b}}{T_\text{fast}}\big)$\\
 \quad at 13.8 m/s$^\text{a}$; $d_2$& $1124 - 976\big( \frac{t-T_{2b}}{T_\text{fast}}\big)$& $1856 - 1237\big(\frac{t-T_{2b}}{T_\text{fast}}\big)$&$2727 - 1079 \big( \frac{t-T_{2b}}{T_\text{fast}}\big)$\\ \bottomrule
\end{tabular}
\raggedright \textsuperscript{a} Taken as average of the range $13.3-14.3$ m/s. \\
\end{table}

\subsection{Race-fit training programs}
\begin{table}[]
    \centering
    \small
\caption{Distances ($d_i$ in metres) covered per week (m/wk) at various speeds for different race-fit training programs undertaken by Thoroughbred racehorses. The distances $d_i$ are ordered from the fastest to the slowest speed. Other phases of training are simulated as described in \autoref{fig:training_program}. Distances covered at 16.7 m/s are assumed to be racing every 14 days (two weeks) over 1600 m races on average, and are unchanged between clusters.}
\label{tab:racefit_properties}
    \begin{tabular}{p{5cm}cccc}
    \toprule
 & \multicolumn{4}{c}{\textbf{Race-fit program group}} \\ \midrule
 & Low volume& Medium volume & Medium-high volume&High volume\\ \midrule
 Distance covered per week [m/wk] ($d_1+d_2+d_3$) & 1733& 2667&3787&4720\\
 \quad at 16.7 m/s; $d_1$& 800& 800& 800&800\\
 \quad at 16.0 m/s; $d_2$& 560& 747&1120&1120\\
 \quad at 13.8 m/s; $d_3$& 373& 1120&
 1867&2800\\ \bottomrule
\end{tabular}
\end{table}
The volume of total distance galloped in race-fit training varies across trainers. The cross-sectional study observed four clusters of workload programs among trainers in Victoria, Australia \citep{MorriceWest2020Mar}. To simulate the impact of different race-fit training programs, the external inputs to the model are adapted to match the clusters of race-fit training programs derived from that study. For conciseness, the `medium-volume program with a higher proportion of high-speed workouts' has been renamed to `medium-high volume program'. A summary of the speeds and distances covered during each training program cluster is given in \autoref{tab:racefit_properties}. All other aspects of the training program are as defined in \autoref{fig:training_program}.

\subsection{Rest duration}
To investigate the impact of rest duration, we simulate two rest periods per year, in line with the average number of rest periods observed in Victoria, Australia \citep{MorriceWest2020Mar}. Thus, the total duration of the training preparation (inclusive of rest) is set to 26 weeks (182 days). A baseline pre-training program of $T_\text{pre-train}=4\ \si{weeks}$ is used, in line with the mean reported in \citet{MorriceWest2020Mar}. The default progressive training program ($T_\text{prog} = 66\ \si{days}$; \autoref{fig:training_program}) is used for these simulations. Of the remaining $88$ days, the rest period $T_\text{rest}$ is varied between 28 to 56 days, and the remaining time is allocated to racing, i.e. $T_\text{race} = 88\ \si{days} - T_\text{rest}$. The distances accumulated at each speed per week are set to their default values in \autoref{fig:training_program}.

\subsection{Rest frequency}

\begin{table}[]
    \centering
    \small
\caption{The duration of rest and training phases ($T$) for training preparation with different rest frequencies for Thoroughbred racehorses.}
\label{tab:rest_frequency}
    \begin{tabular}{lccc}
    \toprule
 & \multicolumn{3}{c}{\textbf{Rests per year}} \\ \midrule
 & 1& 2& 3\\ \midrule
 \textbf{Program duration}& & &\\
 Total program duration; $T_\text{total}$ [weeks]& 52& 26&17.33\\
 Rest duration; $T_\text{rest}$ [weeks]& 8& 6& 4\\
 Pre-training; $T_\text{pre-train}$ [weeks]& 4& 3& 2.33\\
 Progressive training program duration; $T_\text{prog}$ [weeks]& 10& 9&6\\
 Race-fit training program duration; $T_\text{race}$ [weeks]& 30& 8& 5\\  \midrule
 \textbf{Yearly allocation of time}& & &\\
 Rest [weeks]& 8& 12& 12\\
 Pre-training [weeks]& 4& 6 &7\\
 Progressive training [weeks]& 10& 18&18\\
 Race-fit training [weeks]& 30& 16&15\\ \bottomrule
\end{tabular}
\end{table}

The frequency of rest is simulated by varying the number of rests per year $f_\text{rest}$. A summary of the durations of each phase of the training program is given in \autoref{tab:rest_frequency}. With 52 weeks in the year, we set the duration of the entire preparation to $T_\text{total} = 52 / f_\text{rest} \ \si{weeks}$. Following rest, horses undergo pre-training followed by a progressive training period \citet{MorriceWest2020Mar}. The pre-training duration is reduced with the increasing rest frequency due to less reduction in residual fitness when rested over shorter time periods. Since longer rest periods necessitate longer periods of progressive training to restore bone strength \citep{Hitchens2018Jun}, the duration of progressive training decreases with increasing rest frequency. We assume that 40\% of the time in progressive training is allocated to slow work and 60\% of the time is allocated to fast work, in line with the durations defined in \autoref{fig:training_program}. The remaining time within the entire training program is allocated to the race-fit training (i.e., $T_\text{race}=T_\text{total}- T_\text{rest}-T_\text{pre-train}-T_\text{prog}$). The times in progressive and race-fit training vary with rest frequency, with lower rest frequencies resulting in more weeks available for racing. Training preparations with two and three rests per year allow 16 and 15 weeks of racing, respectively. As we assume one race every two weeks (14-day period), this corresponds to approximately eight races per year, which falls within the typical ranges reported \citep{morrice2025linkage}. In contrast, programs with one rest per year accommodate 15 races due to the longer time within the race-fit training phase, and thus more time available for racing.

\subsection{Low-intensity training}
To investigate the effect of substituting low intensity training (colloquially also known as ``backing off") for periods of extended rest, we define a new training program that starts with a rest period of $T_\text{rest} = 6\ \si{weeks}$, followed by 4 weeks of pre-training and a progressive training program of length $T_\text{prog} = 10\ \si{ weeks}$. For progressive training programs, we allocate 40\% of the time to the slow phase and 60\% of the time to the fast phase, in line with their proportions in \autoref{fig:training_program}. Following progressive training, the horse undertakes four periods of race-fit training interspersed by three periods of low-intensity training. We simulate the outcomes of two types of low-intensity training: a longer four-week break from racing, and a shorter two-week break.

\subsubsection{Four-week back-off}
Within the low-intensity training with four-week back-off, the duration of low-intensity training lasts for $T_\text{low} = 4\ \si{weeks}$ (12 weeks over a year). The horse spends the first two of these weeks trotting and/or cantering (6 m/s) for 2000m/day. Following this, to progress to race-level fitness, the horse is trained at a canter (7.5 m/s) for 2000 m/day for one week, and then spends another week at three-quarter pace (11.8 m/s) for 257 m/day. To achieve a program duration of 52 weeks, the duration of each racing campaign is set to $T_\text{race} = 5\ \si{weeks}$. As the horse spends the first two weeks trotting  and/or cantering at 6 m/s and the third week cantering at 7.5 m/s, the default slow workouts (as described in \autoref{fig:training_program}, purple box) are paused during the first 3 weeks, but resume once the horse trains at gallop speeds and above.

\subsubsection{Two-week back-off}
When a two-week back-off is considered, the periods of low-intensity training last for $T_\text{low} = 2\ \si{weeks}$ (6 weeks over a year). During this time, the training is reduced to three-quarter pace (11.8 m/s) for 257 m/day. Following this, the horse immediately returns to a race-fit training program. To achieve a program duration of 52 weeks,, the duration of each racing campaign is set to $T_\text{race} = 6.5\ \si{weeks}$. Slow workouts are assumed to continue during this break.

\section{Results}
\subsection{Progressive training}
\begin{figure}
    \centering
    \includegraphics[width=\linewidth]{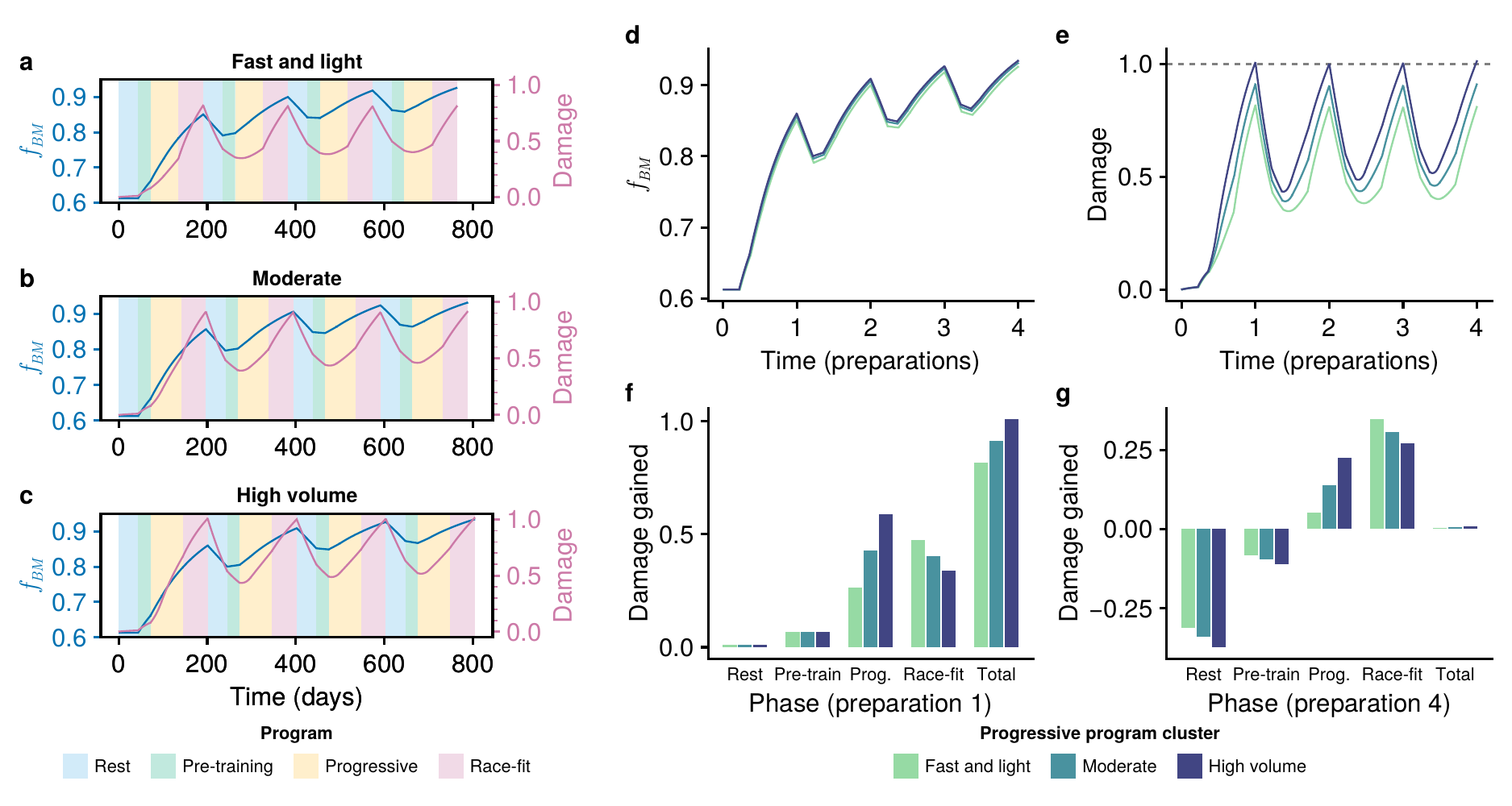}
    \caption{A comparison of progressive training programs. (a--c) Simulation of bone volume fraction (blue lines) and damage (pink lines) for training programs incorporating (a) fast and light; (b) moderate; and (c) high volume progressive training preparations, as defined by \citet{MorriceWest2020Mar}. The damage $D^*$ is a normalised variable, where bone failure occurs at $D^*=1$. Backgrounds are coloured based on the phase of training during preparation. (d--e) A comparison of the evolution of (d) bone volume fraction and (e) bone damage for different progressive training programs. The grey dashed line indicates the threshold at which the bone is considered to have failed. (f--g) A comparison of net changes in damage over different phases of preparation. (f) First preparation; (g) final (fourth) preparation. Changes are defined as the difference between the final value relative to the initial value within a phase of training preparation.}
    \label{fig:progressive_program_comparison}
\end{figure}
Results from simulating the response of bone to different progressive training program clusters are shown in \autoref{fig:progressive_program_comparison}. We consider a total of four training preparations, from an initial bone volume corresponding to an untrained racehorse with no bone damage (\autoref{fig:progressive_program_comparison}a--c). During the rest phase, both bone volume fraction and damage decrease, while these values increase as a result of progressive training. The choice of progressive program has very little effect on the bone volume fraction which is predominantly driven by differences in the number of days in progressive training. However, there is a noticeable effect of progressive program cluster on bone damage, with the high volume cluster (\autoref{fig:progressive_program_comparison}c) resulting in greater bone damage at the end of four preparations in comparison to the fast and light cluster (\autoref{fig:progressive_program_comparison}a).

A comparison of the response of bone volume fraction to different progressive training programs is summarised in \autoref{fig:progressive_program_comparison}d, with time normalised to the number of preparations (noting that the total duration of the preparation is different between clusters). A gradual increase in the volume fraction with the increasing number of preparations is observed, with the greatest gain seen in the first preparation. This is expected as the unadapted bone is exposed to loading during the progressive phase in the first training preparation. After the first preparation, the bone is already adapted to some level of training, and the rate of adaptation is slowed. However, the overall effect of different progressive programs on bone volume fraction is minor, reflecting the simulations in \autoref{fig:progressive_program_comparison}a--c.

\autoref{fig:progressive_program_comparison}e illustrates a comparison of bone damage between progressive training clusters. In contrast to bone volume fraction, the minimum and maximum damage appear to stabilise within one preparation. In addition, there are more substantial variations in bone damage between training clusters, with the damage increasing with the volume of training. In particular, the maximum damage observed at the end of race-fit training (1.016) is higher for the high volume cluster relative to the fast and light cluster (0.813) (\autoref{tab:progressive_results}). Even at the end of the first preparation, the maximum damage observed in the high volume cluster crosses the threshold ($D^*=1$, dashed horizontal line), which corresponds to the failure of a local volume of bone within the lateral condyle of the third metacarpal bone through excessive accumulation of microdamage. This does not necessarily lead to a gross fracture of bone, but may result in other bone injuries or pathology such as palmar osteochondral disease (POD), which is commonly observed in active racehorses  \citep{pinchbeck2013pathological}.

The change in damage across the first preparation (\autoref{fig:progressive_program_comparison}f) and final preparation (\autoref{fig:progressive_program_comparison}g) is plotted, and subdivided into contributions by training phase. During the first preparation (starting at zero damage), all phases result in net accumulation of damage, with the damage in progressive training greater in high volume programs relative to lower volume programs. Interestingly, despite having implemented a uniform race-fit phase across the progressive training cluster (i.e., identical training load and duration between clusters), the net accumulation of damage in the race-fit phase is lower for higher-volume programs (fourth column, \autoref{fig:progressive_program_comparison}f). This effect is because horses with higher volumes of progressive training have more damage at the end of the progressive phase (third column, \autoref{fig:progressive_program_comparison}f). That is, horses commence the race-fit training phase with non-identical pre-existing damage, and thereby, higher damage increases the rate of damage repair. Despite this, higher-volume programs still accumulate greater damage over the entire preparation (\autoref{fig:progressive_program_comparison}f; right column).

The damage accumulated over the final (fourth) preparation provides an indication of the contributions of phases of training following the stabilisation of damage in response to the first few preparations  (\autoref{fig:progressive_program_comparison}g). Due to the presence of pre-existing damage, both the rest and pre-training programs result in the net repair of damage, with the repair being greater for high-volume programs where the extent of bone damage is greater. As with the first preparation, damage accumulated during progressive training increases with training volume, and damage accumulated during the race-fit training decreases with progressive training volume. Unlike the first preparation, the majority of damage accumulated occurs during race-fit training rather than progressive training, reflecting the higher volumes of gallops (and hence the rate of damage formation) during race-fit training. A summary of the bone volume fraction, bone damage and distance accumulated over the final (fourth) preparation is provided in \autoref{tab:progressive_results}.

\begin{table}
\small
\caption{Summary of simulations for different progressive training programs in Thoroughbred racehorses.}
\centering \begin{tabular}{lccc}
    \toprule
    \textbf{Progressive program} & \textbf{Fast and light} & \textbf{Moderate} & \textbf{High volume}\\
    \midrule
    Minimum $f_{BM}$$^\text{a}$ & 0.858 & 0.864 & 0.867\\
    Maximum $f_{BM}$$^\text{a}$ & 0.926 & 0.931 & 0.935\\
    Minimum damage$^\text{a}$ & 0.401 & 0.460 & 0.517\\
    Maximum damage$^\text{a}$ & 0.813 & 0.913 & 1.016\\
    Distance covered (km)$^\text{b}$ &  &  & \\
    $\quad$ Canter (7.5 m/s) & 126.0 & 138.0 & 146.0\\
    $\quad$ Pace work to fast gallop  (11.8--16 m/s) & 8.9 & 16.6 & 26.0\\
    Damage formed $^\text{b}$ & 0.236 & 0.374 & 0.512\\
    Damage repaired $^\text{b}$ & 0.184 & 0.236 & 0.288\\
    \bottomrule
\end{tabular}\\
\raggedright \textsuperscript{a}Values summarised over the final (fourth) preparation, including rest, pre-training, progressive training and race-fit training. \\
\raggedright \textsuperscript{b}Values summarised over the final (fourth) progressive phase. \\
\label{tab:progressive_results}
\end{table}

\subsection{Race-fit training}
\begin{figure}
    \centering
    \includegraphics[width=\linewidth]{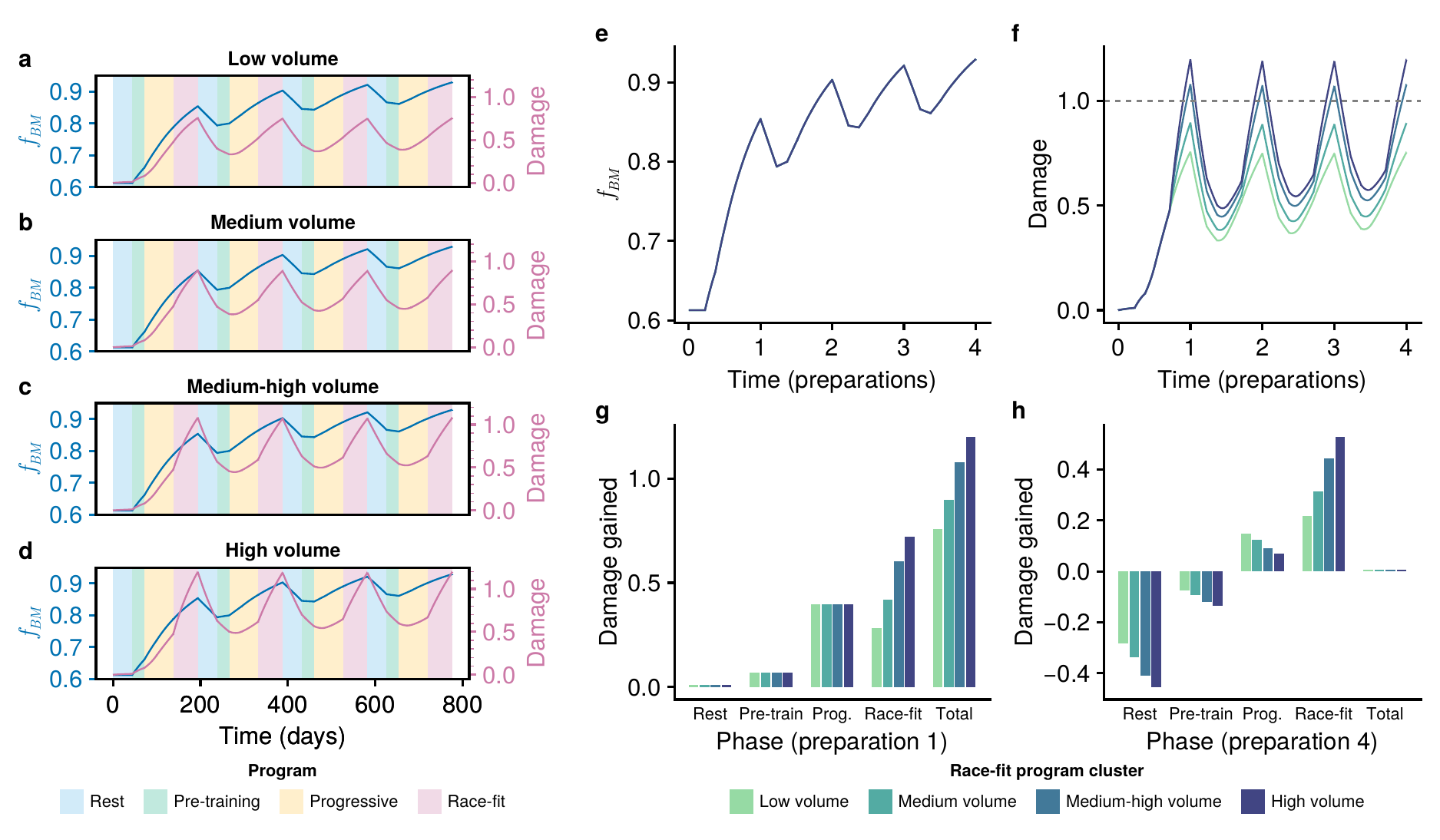}
    \caption{A comparison of race-fit training programs. (a--d) Simulation of bone volume fraction (blue lines) and damage (pink lines) for training programs incorporating (a) low; (b) medium; (c) medium-high and (d) high volume race-fit training program, as defined by \citet{MorriceWest2020Mar}. The damage $D^*$ is a normalised variable, where bone failure occurs at $D^*=1$. Backgrounds are coloured based on the phase of training preparation. (e--f) A comparison of the evolution of (e) bone volume fraction and (f) bone damage for different race-fit training programs.  The grey dashed line indicates the threshold at which the bone is considered to have failed. (g--h) A comparison of changes in damage over different phases of a preparation. (g) First preparation; (h) final (fourth) preparation. Changes are defined as the difference in final value relative to the initial value within a phase of training.}
    \label{fig:racefit_program_comparison}
\end{figure}

We next investigate the impact of varying race-fit training volumes on bone adaptation and damage accumulation, whilst keeping the remaining phases (i.e., rest, pre-training and progressive training) identical. Simulations of these training programs under varying race-fit program volumes are shown in \autoref{fig:racefit_program_comparison}a--d. As with progressive training, increasing the volume of race-fit training leads to a negligible difference in bone volume fraction and greater accumulation of bone damage. The model predicts bone failure by the end of each racing preparation for both the medium-high volume and high volume cluster (\autoref{fig:racefit_program_comparison}f, \autoref{tab:racefit_summary}).

The damage recorded during each preparation is shown in \autoref{fig:racefit_program_comparison}f. Similar to changes in progressive training, increased race-fit training volumes lead to greater damage. The effects are particularly pronounced for the maximum damage observed during a preparation, which is beyond the bone's fatigue life for higher volumes of race-fit training. In general, variations in race-fit training programs appear to lead to greater effects on bone damage when compared to progressive training, with the final damage following four preparations varying from 0.753 to 1.196 (\autoref{tab:racefit_summary}).

A breakdown of the damage accumulated in each phase during the first and final preparations is given in \autoref{fig:racefit_program_comparison}g--h. During the first preparation, damage accumulation is uniform until the commencement of the race-fit training phase (\autoref{fig:racefit_program_comparison}g). This is due to keeping all the other phases of the preparation identical to test the effects of race-fit training only. Following progressive training during the first preparation, high volume race-fit training leads to greater damage accumulated over the entire training preparation (\autoref{fig:racefit_program_comparison}g). Over the final preparation, there is a similar trend of total race-fit phase damage increasing with the volume of training (fourth column, \autoref{fig:racefit_program_comparison}h). However, the net increase in damage accumulated is lower compared to the first preparation due to the greater repair rates stimulated by the pre-existing damage (fifth column, \autoref{fig:racefit_program_comparison}h). Similarly, because of the pre-existing damage, repair during rest and pre-training is enhanced for high-volume programs, and the net damage accumulated during the progressive phase is reduced for high-volume clusters (third column, \autoref{fig:racefit_program_comparison}h). A summary of the bone volume fraction, bone damage and distance accumulated over the final (fourth) preparation is provided in \autoref{tab:racefit_summary}.

\begin{table}
\small
\caption{Summary of simulations for different race-fit training programs of Thoroughbred racehorses.}
\centering \begin{tabular}{lcccc}
    \toprule
    \textbf{Race-fit program} & \textbf{Low volume} & \textbf{Medium volume} & \textbf{Medium-high volume} & \textbf{High volume}\\
    \midrule
    Minimum $f_{BM}$$^\text{a}$ & 0.861 & 0.861 & 0.861 & 0.861\\
    Maximum $f_{BM}$$^\text{a}$ & 0.929 & 0.929 & 0.929 & 0.929\\
    Minimum damage$^\text{a}$ & 0.387 & 0.447 & 0.525 & 0.574\\
    Maximum damage$^\text{a}$ & 0.753 & 0.893 & 1.077 & 1.196\\
    Distance covered (km)$^\text{b}$ &  &  &  & \\
    $\quad$ Canter (7.5 m/s) & 111.8 & 111.8 & 111.8 & 111.8\\
    $\quad$ Pace work to fast gallop  (11.8--16 m/s) & 7.5 &  14.9 & 23.9 & 31.3\\
    Damage formed $^\text{b}$ & 0.418 & 0.545 & 0.712 & 0.82\\
    Damage repaired $^\text{b}$ & 0.203 & 0.232 & 0.27 & 0.294\\
    \bottomrule
\end{tabular}
\raggedright \textsuperscript{a}Values summarised over the final (fourth) preparation, including rest, pre-training, progressive training and race-fit training. \\
\raggedright \textsuperscript{b}Values summarised over the final (fourth) race-fit phase. \\

\label{tab:racefit_summary}
\end{table}

\subsection{Rest duration}
\begin{figure}
    \centering
    \includegraphics[width=\linewidth]{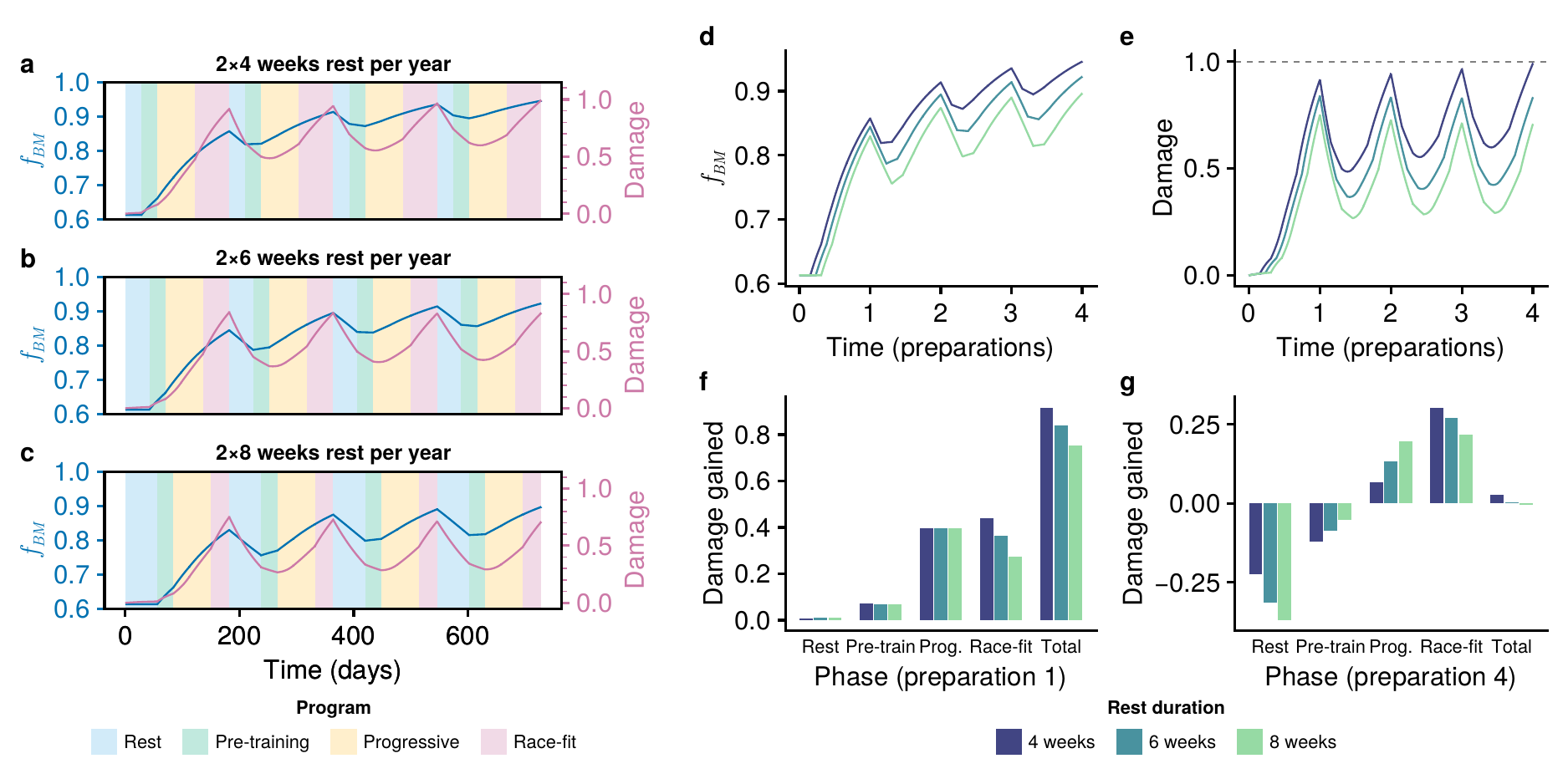}
    \caption{The effect of rest duration on damage accumulated during training. (a--c) Simulations of racehorse training programs are run with rest durations of (a) 4 weeks; (b) 6 weeks; and (c) 8 weeks. Horses are rested two times per year. The damage $D^*$ is a normalised variable, where bone failure occurs at $D^*=1$. Backgrounds are coloured based on the phase of training during a preparation. (d--e) A comparison of the evolution of (d) bone volume fraction and (e) bone damage for different rest durations.  The grey dashed line indicates the threshold at which the bone is considered to have failed. (f--g) A comparison of changes in damage over different phases of a preparation. (f) First preparation; (g) final (fourth) preparation. Changes are defined as the difference in final value relative to the initial value within a phase of training.}
    \label{fig:rest_duration_summary}
\end{figure}
Simulations of training programs with varying rest durations (4, 6 and 8 weeks) are shown in \autoref{fig:rest_duration_summary}a--c, keeping the number of rests per year static (2 rests per year), which results in a duration of 26 weeks for each training preparation. Due to increased time in rest and reduced time in racing, the bone volume fraction is lower for horses with longer rest durations. The highest damage is observed in horses with rest durations of 4 weeks, where the bone is near failure (0.992) by the end of four preparations. The bone damage decreases with increased rest, down to 0.713 after four preparations with a rest duration of 8 weeks (\autoref{tab:rest_durations}). These trends are also shown in  \autoref{fig:rest_duration_summary}d--e, which records the ranges of bone volume fraction and damage observed during the final preparation for rest durations varying from 4 to 8 weeks. The duration of rest has a particularly marked impact on the minimum damage achieved during the final preparation, which ranges from 0.292 (8 weeks rest) to 0.599 (4 weeks rest).

\begin{table}
\small
\caption{Summary of simulations for different rest durations between training and racing preparations of Thoroughbred racehorses over one year.}
\centering \begin{tabular}{lccc}
    \toprule
    \textbf{Rest duration} & \textbf{4 weeks} & \textbf{6 weeks} & \textbf{8 weeks}\\
    \midrule
    Minimum $f_{BM}$$^\text{a}$ & 0.895 & 0.856 & 0.814\\
    Maximum $f_{BM}$$^\text{a}$ & 0.946 & 0.922 & 0.896\\
    Minimum damage$^\text{a}$ & 0.599 & 0.423 & 0.292\\
    Maximum damage$^\text{a}$ & 0.992 & 0.833 & 0.713\\
    \midrule
     \textbf{Yearly allocation of time}& & &\\
     Rest [weeks]& 8& 12&16\\
     Pre-training [weeks] & 8 & 8 & 8 \\
     Progressive training [weeks]& 18.9& 18.9&18.9\\
     Racing [weeks]& 17.1& 13.1&9.1\\ \bottomrule
\end{tabular}\\
\raggedright \textsuperscript{a}Values summarised over the final (fourth) preparation, including rest, progressive training and race-fit training.
\label{tab:rest_durations}
\end{table}

The contributions of various phases of training to bone damage are given in \autoref{fig:rest_duration_summary}f--g. During the first preparation (\autoref{fig:rest_duration_summary}f), the damage accumulated is identical until the race-fit phase, where the damage accumulated from racing is lower for longer rest durations due to a shorter duration available for racing preparation. For the final preparation (\autoref{fig:rest_duration_summary}g), the damage removed by rest is increased for longer rest periods. However, the net damage accumulated by progressive training increases with rest duration. Consistent with the first preparation, the damage accumulated during the race-fit training decreases slightly with rest duration, although to a lesser extent than in the first preparation.

\subsection{Rest frequency}
\begin{figure}
    \centering
    \includegraphics[width=\linewidth]{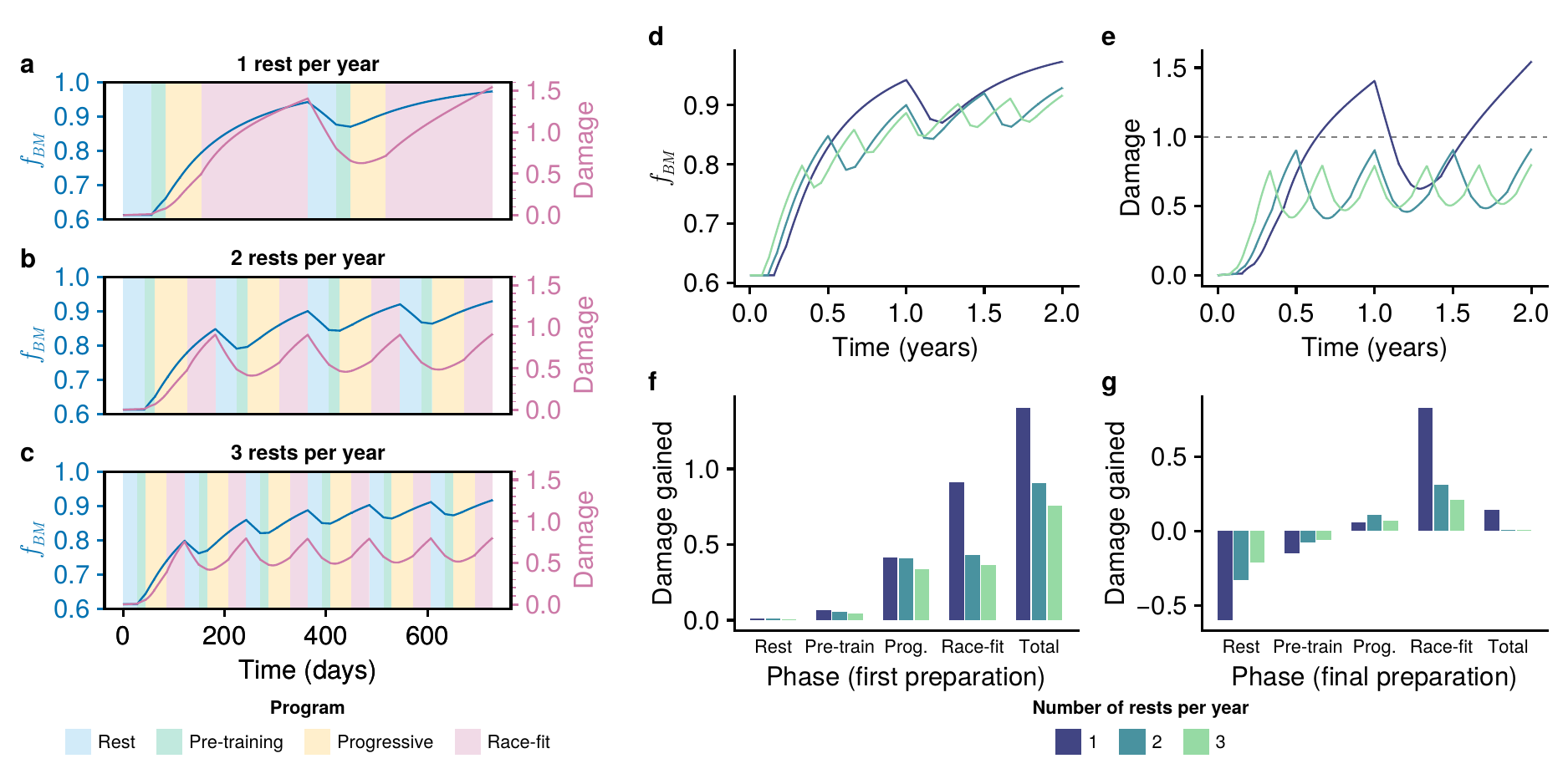}
    \caption{The effect of rest frequency on damage accumulated during training. (a--c) Simulation of bone volume fraction (blue lines) and damage (pink lines) for training programs with (a) 1 rest per year; (b) 2 rests per year; and (c) 3 rests per year. The damage $D^*$ is a normalised variable, where bone failure occurs at $D^*=1$. Backgrounds are coloured based on the phase of training during a preparation. (d--e) A comparison of the evolution of (d) bone volume fraction and (e) bone damage for different rest frequencies.  The grey dashed line indicates the threshold at which the bone is considered to have failed. (f--g) A comparison of changes in damage over different phases of a preparation. (f) First preparation; (g) final preparation, preceding the end of the second year. Changes are defined as the difference in final value relative to the initial value within a phase of training.}
    \label{fig:rest_frequency_summary}
\end{figure}

\begin{table}
\small
\caption{Summary of simulations for different rest frequencies over one year of training and racing preparations of Thoroughbred racehorses.}
\centering \begin{tabular}{lccc}
    \toprule
    \textbf{Rests per year} & \textbf{1} & \textbf{2} & \textbf{3}\\
    \midrule
    Minimum $f_{BM}$$^\text{a}$ & 0.87 & 0.843 & 0.848\\
Maximum $f_{BM}$$^\text{a}$ & 0.973 & 0.929 & 0.916\\
Minimum damage$^\text{a}$ & 0.626 & 0.459 & 0.49\\
Maximum damage$^\text{a}$ & 1.543 & 0.911 & 0.799\\
    \bottomrule
\end{tabular}\\
\raggedright \textsuperscript{a}Values summarised over the final (second) year, including rest, progressive training and race-fit training.
\label{tab:rest_frequency_results}
\end{table}

The response of bone to different rest frequencies per year is presented in \autoref{fig:rest_frequency_summary}a--c. Because lower rest frequencies result in longer durations in race-fit training, the bone volume fraction and damage exhibit greater fluctuations in comparison to more frequent rest. Programs with more frequent rest result in lower maximum bone volume fractions, although the minimum bone volume fraction is relatively similar between groups, particularly with two and three rest frequencies (\autoref{fig:rest_frequency_summary}d, \autoref{tab:rest_frequency_results}). Notably, horses with one rest per year reach far greater maximum damage (1.543) compared to horses that rest twice per year (0.911) (\autoref{fig:rest_frequency_summary}e), with further reductions in maximum bone damage achieved by increasing to three rests per year (0.799). Summaries of damage accumulated during phases of training are shown in \autoref{fig:rest_frequency_summary}f--g, highlighting the greater fluctuations in damage seen in programs with one rest per year.

\subsection{Low-intensity training}
\begin{figure}
    \centering
    \includegraphics[width=\linewidth]{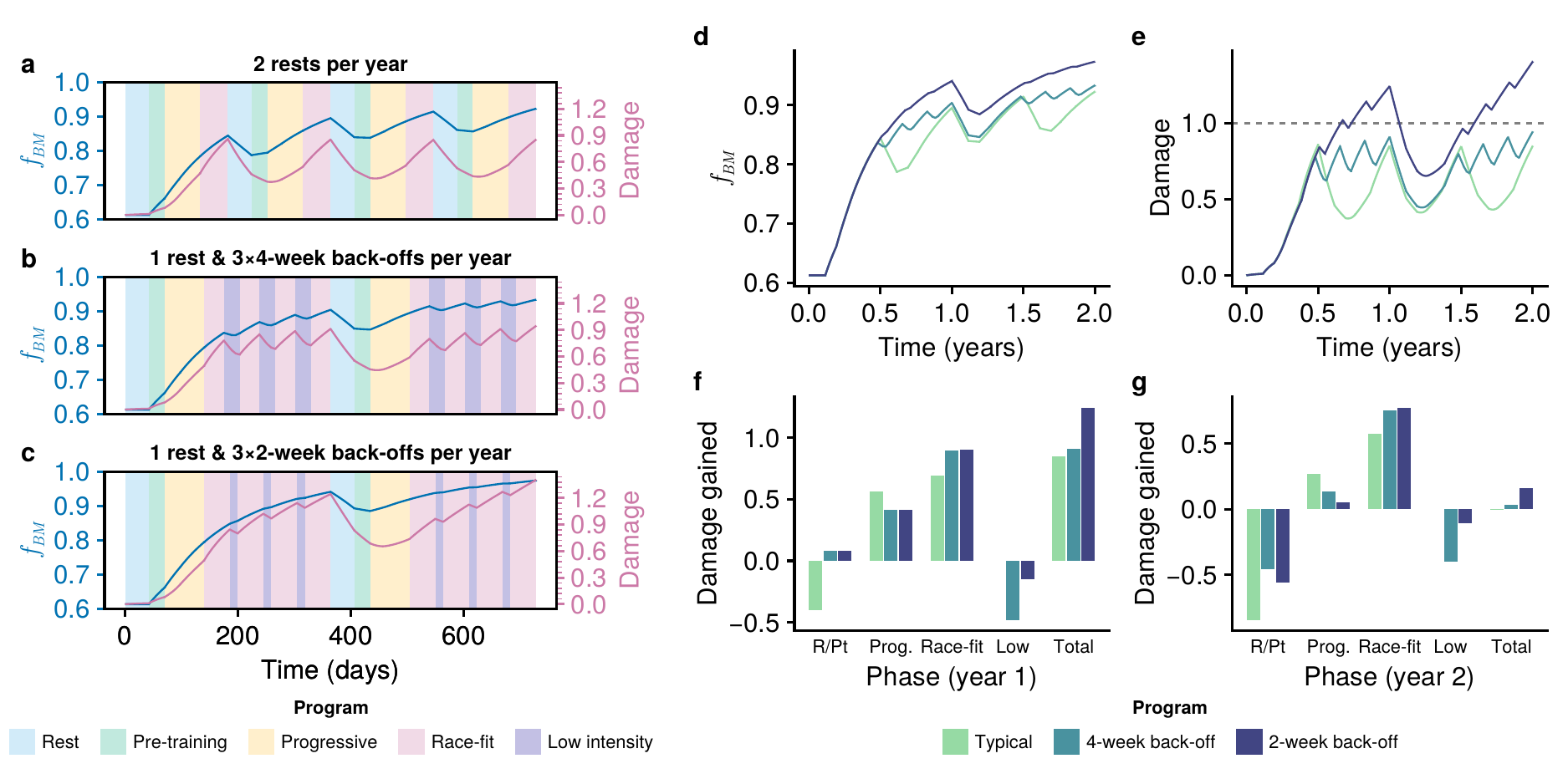}
    \caption{The effect of low-intensity training on damage accumulated during training. (a--c) Simulation of bone volume fraction (blue lines) and damage (pink lines) for training programs with (a) a typical rest program with two rests (6 weeks in duration) per year; (b) a program with one rest per year, with racing preparations interspersed with three four-week periods of low intensity training where workload is reduced to trot (3.5 m/s) before working back up to galloping; and (c) a program with one rest per year, with racing preparations interspersed with three two-week periods of low intensity training at slow gallop (11.8 m/s). The damage $D^*$ is a normalised variable, where bone failure occurs at $D^*=1$. Backgrounds are coloured based on the phase of training during a preparation. (d--e) A comparison of the evolution of (d) bone volume fraction and (e) bone damage for each program.  The grey dashed line indicates the threshold at which the bone is considered to have failed. (f--g) A comparison of changes in damage over different phases of a preparation. (f) First year; (g) second year. Changes are defined as the final value relative to the initial value within a phase of training, and are summed in cases where certain phases occur multiple times during a year. The bars corresponding to R/Pt are the combined contributions of rest and pre-training.}
    \label{fig:freshenup_summary}
\end{figure}

\begin{table}
\small
\caption{Summary of simulations for low intensity training of Thoroughbred racehorses.}
\centering \begin{tabular}{lccc}
    \toprule
    \textbf{Program} & \textbf{ Typical (2 rests per year)} & \textbf{4-week back-off} & \textbf{2-week back-off}\\
    \midrule
Minimum $f_{BM}$$^\text{a}$ & 0.843 & 0.846 & 0.885\\
Maximum $f_{BM}$$^\text{a}$ & 0.929 & 0.933 & 0.973\\
Minimum damage$^\text{a}$ & 0.459 & 0.446 & 0.654\\
Maximum damage$^\text{a}$ & 0.911 & 0.946 & 1.406\\
    \midrule
     \textbf{Yearly allocation of time}& & &\\
     Rest [weeks]& 12& 6&6\\
     Pre-training [weeks] & 6 & 4 & 4 \\
     Progressive training [weeks]& 18& 10&10\\
     Racing [weeks]& 16& 20&26\\ 
     Low intensity training [weeks]& 0& 12& 6\\ \bottomrule
\end{tabular}\\
\raggedright \textsuperscript{a}Values summarised over the final (second) year, including rest, progressive training and race-fit training.
\label{tab:freshenups}
\end{table}

The effects of introducing low-intensity training as an alternative to rest are shown in \autoref{fig:freshenup_summary}a--c. In comparison to a typical training program with two rests per year (\autoref{fig:freshenup_summary}a, identical to program in \autoref{fig:rest_frequency_summary}b), programs with rests replaced by 4-week periods of low-intensity training in between race-fit training result in similar values for maximum damage, but with smaller fluctuations in both damage and bone volume fraction (\autoref{fig:freshenup_summary}b,e). However, replacing rests with 2-week periods of low-intensity training is insufficient to mitigate bone damage, leading to bone failure at the end of each preparation (\autoref{fig:freshenup_summary}c). \autoref{fig:freshenup_summary}d--e emphasise both the increased bone volume fraction and damage with the replacement of rests with low-intensity training of duration 2 weeks. During both the first and second years, the introduction of low-intensity training in place of rest reduces the damage repaired by rest and pre-training (fourth column \autoref{fig:freshenup_summary}f--g). While three bouts of 4-week low intensity training are able to compensate for this difference, this is not the case when low-intensity periods are reduced to 2 weeks, as indicated by the relatively small repair rates observed in the low intensity phase of \autoref{fig:freshenup_summary}g (fourth column).

\begin{figure}
    \centering
    \includegraphics[width=.8\linewidth]{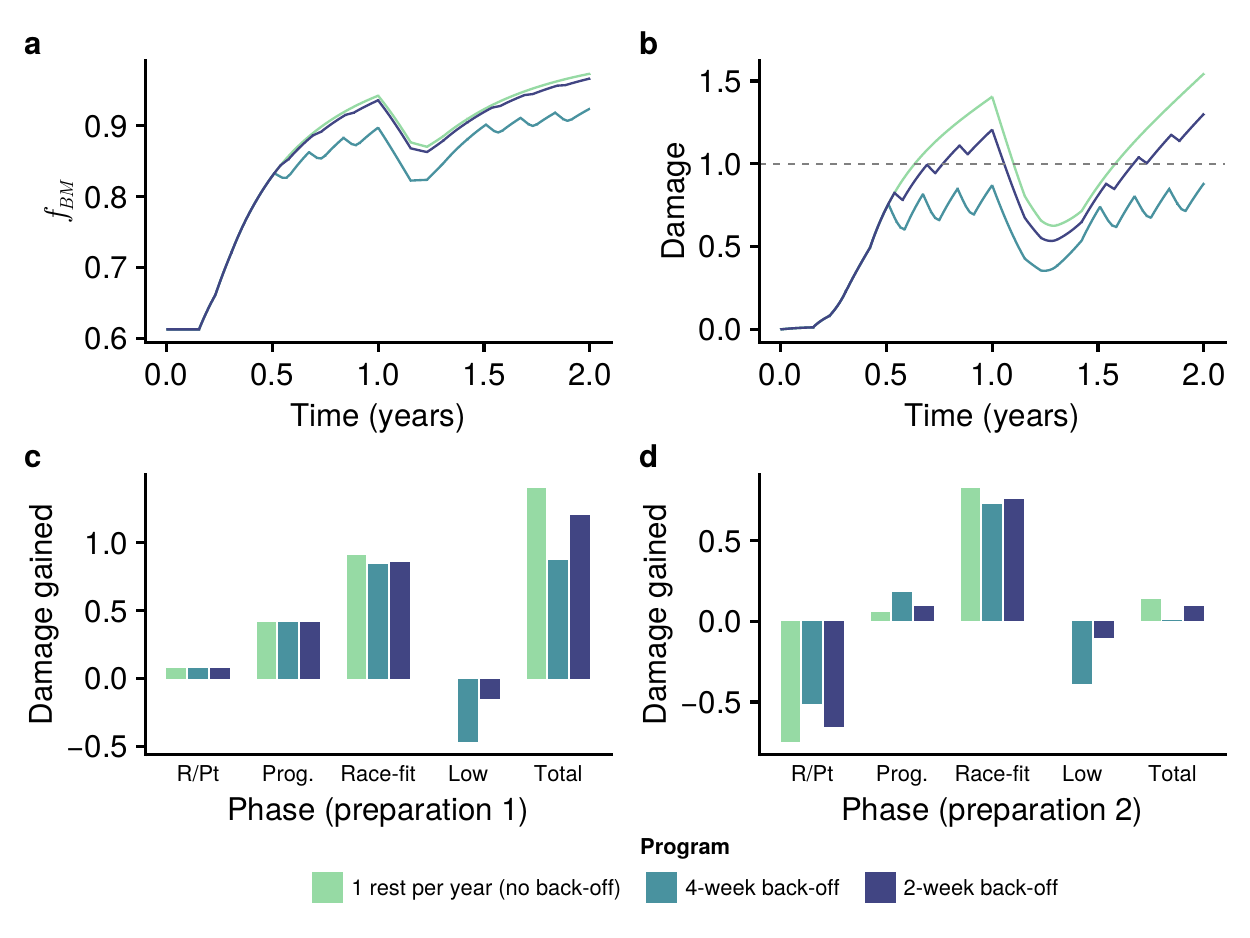}
    \caption{The effect of low-intensity training on damage accumulated during training compared to a training program with one rest per year. (a--b) A comparison of the evolution of (a) bone volume fraction and (b) bone damage for each program. The damage $D^*$ is a normalised variable, where bone failure occurs at $D^*=1$ (grey horizontal dashed line). (c--d) A comparison of changes in damage over different phases of a preparation (first (c) and second year (g)). Changes are defined as the difference in final value relative to the initial value within a phase of training, and are summed in cases where certain phases occur multiple times during a year. The bars corresponding to R/Pt are the combined contributions of rest and pre-training.}
    \label{fig:freshenup_summary_2}
\end{figure}

To assess the benefit of low-intensity training, we also compare the observed bone volume fractions and damage with those seen in training programs with one rest per year (\autoref{tab:rest_frequency}).  With one rest per year, horses spend 30 weeks in the race-fit phase during each preparation, leading to greater bone damage. The damage incurred during the prolonged race-fit duration can be substantially reduced (from a maximum of 1.543 to 0.946) by introducing low-intensity training with a 4-week back-off (\autoref{fig:freshenup_summary_2}b), although the benefit is less apparent when 2-week back-offs are introduced. The extent of bone adaptation is lower in horses that have 4-week back-off compared to horses with 2-week or no back-off (\autoref{fig:freshenup_summary_2}a), which may account for the higher resorption rates (and hence rates of damage repair) seen in that program (4-week back-off) compared to others (2-week or no back-off). The evolution of bone volume fraction is similar between programs with one rest per year (no back-off) and a 2-week back-off (\autoref{fig:freshenup_summary_2}a).

\section{Discussion}

The training program prescribed to Thoroughbred racehorses before participating in races is known to influence the risk of musculoskeletal injury \citep{Wong2023Association}. However, it is often difficult to uncouple the effects of training programs from other sources of variation in observational studies. In this study, we use mathematical modelling to estimate bone dmage by incorporating biological processes and the response of bone to training on a continuous scale. We extended our previous mathematical model of bone adaptation and damage \citep{Pan2025Apr} to investigate the effect of different training strategies on bone damage. The model coupled the processes of bone adaptation
and damage in the metacarpal subchondral bone of Thoroughbred racehorses and was calibrated to bone morphology data of racehorses in training and rest. In this work, we explored the effects of different training volumes and rest strategies derived from trainer surveys conducted in Victoria, Australia \citep{MorriceWest2020Mar} on bone adaptation and damage. Our key findings are that lower volumes of progressive and race-fit training reduce damage accumulation. Similarly, longer and more frequent rest periods reduce damage accumulation but also reduce the time available for racing (Figures \ref{fig:rest_duration_summary}--\ref{fig:rest_frequency_summary}), whereas multiple 4-week periods of lower volume training reduce damage accumulation similar to that of a 8-week rest period while increasing the time available for racing (\autoref{fig:freshenup_summary}-\ref{fig:freshenup_summary_2}). 

According to our simulations, increasing training volume leads to greater bone damage accumulation over the course of a training preparation. This is generally consistent with other studies, for example, showing that fracture risk increases particularly with greater cumulative distance at canter and gallop over a one-month period \citep{Verheyen2006Dec}. Varying the volumes of both progressive and race-fit training led to substantial changes in the damage accumulated after four preparations (Figures \ref{fig:progressive_program_comparison}--\ref{fig:racefit_program_comparison}). We find that the volume of race-fit training has a greater impact on damage accumulation than progressive training, possibly reflecting the larger high-speed training and racing volumes observed in such programs. This is consistent with meta-analyses that increased high-speed exercise distance increased the risk of MSI \citep{crawford2020effect,crawford2021risk} and FMI \citep{hitchens2019meta}. The results indicate the importance of monitoring gallop distances in both progressive and race-fit training (in addition to racing) in assessing the risk of injury.  The type of progressive or race-fit programs had little impact on the bone volume fractions observed. Progressive training programs play a crucial role in adapting bone to the stresses seen in races \citep{evans1994training}, but we found little evidence that increasing the duration or volume of progressive training reduced the risk of bone injury. While there is strong evidence that the fatigue life of bone is related to the bone volume fraction \citep{martig2020relationship}, there is still some uncertainty on the mathematical relationship between these variables due to individual variation, which we did not account for in this study.

Our findings suggest that increasing rest had a positive impact on preventing bone injury. This is in line with the hypothesis that extended or more frequent rest periods are likely to be beneficial in repairing bone damage \citep{Holmes2014Dec,Pan2025Apr}. Increasing rest duration resulted in greater bone damage repair (during rest), although some of this benefit is lost once the horse has trained for a period of time. Nonetheless, even providing horses with two weeks of additional rest (from six to eight weeks) can have substantial benefits for damage repair, as the maximum damage was reduced by 14.4\% (from 0.833 to 0.713). The frequency of rest also had a large impact on the damage incurred. In particular, using typical training volumes in Australia, resting horses once per year (compared to twice per year) results in the maximum recorded damage increasing from 0.852 to 1.543, reflecting a marked increased risk of injury. This increase in damage was largely a consequence of participation in more races and the extended periods available for racing once the horse was race-fit. Furthermore, to assess the effects of training, we fixed race distances to a middle distance race of 1,600 meters (approximately one mile). Hence, horses that race over longer distances (e.g. 3,200 meters) would be expected to incur much greater bone damage. We therefore suggest that it is potentially risky to rest horses only once per year without other adjustments to the training program to mitigate bone damage accumulation. While further benefits are seen when increasing the frequency of rest from two to three times per year, the benefit was less marked, and such programs result in fewer weeks available for active racing. In addition, there needs to be careful consideration when implementing changes in rest practices because horses are at higher risk of injury after a period of rest, as the bone adaptation dissipates during a period of rest and requires subsequent re-training for recovery of those adaptations \citep{Holmes2014Dec,carrier1998association,Wong2023Association}.

We investigated the effect of replacing one rest period with several periods of low-intensity training over the course of the year. Such a strategy would allow horses to race for more weeks per year, but little is known about the potential impacts on the risk of bone injury. While our simulations only modelled one level of maintenance and progressive training intensity, they suggest that replacing rest with low-intensity training could potentially be a viable option to reduce bone injury risk. However, the strategy is only effective when back-offs are long enough (4 weeks), and workloads are adequately reduced to trotting and cantering for the first few weeks, before introducing more intense gallops (\autoref{fig:freshenup_summary}). This strategy may also be less effective in certain horses due to variations in joint loading at a given speed that may arise from other factors, such as, combined horse and jockey weight and horse gait \citep{merritt2008influence}.  We found that backing off horses for two weeks was an ineffective strategy for reducing bone damage. Though resting horses once a year is less common, in jurisdictions that more routinely prescribe one rest per year (UK, New Zealand), horses tend to have longer periods between races and lower training volumes \citep{verheyen2005training,perkins2005profiling}; other training programs employed globally are worth investigating in the future.

We acknowledge that many of the training strategies that reduce the risk of injury also tend to lead to less opportunity for racing, and that trainers may be reluctant to provide horses with more rest if this results in less earnings. However, lower volumes of high-speed work during progressive and race-fit training do not affect the time available for racing, and prior studies found no significant relationship between training volumes and trainer success \citep{MorriceWest2021Nov}. In addition, we postulate that these strategies to maintain the horse's bone health would increase career longevity, which would also result in greater opportunities to race over more preparations and years. 

As with all modelling studies, our model also comes with a number of limitations. While the effect of training volume (strides per day) is simulated to induce bone damage in our model, we have not included the effect of strides per day on bone formation and resorption rate. Furthermore, we have not included variations in either the biological rates of horses or variations in training regimens (within clusters). Incorporating such heterogeneities would enable better estimates of bone injury risk and potentially allow models to be better tailored to individuals. However, we expect that the trends seen between simulations would still hold in spite of this variation. In addition, we have not accounted for impairment of bone repair once bone damage passes a threshold that disrupts vasculature \citep{saran2014role}. This will result in reduced repair rather than the higher rates of repair the model predicts for the higher impact training regimens. Furthermore, even though the damage passes the threshold ($D^*=1$) defined in our model, which corresponds to the localised failure of a volume of bone, we assume that the horse continues training. This might not happen in practice. In addition, our results indicate that race-fit training programs with medium to high volume workloads are associated with an increased risk of injury. However, these programs do not appear to produce clearly observable or clinically evident bone injury as, such training approaches remain widely used by trainers in practice. One possible explanation is that more resilient horses are preferentially able to tolerate these programs, or alternatively, that a substantial burden of underlying bone injury remains undetected. Furthermore, cardiovascular fitness variables, such as maximal oxygen uptake and heart rate, are known to increase with structured high-speed training but decline within weeks of reduced workload \citep{evans2007physiology,de2019cardiovascular}. We have not investigated the effect of training on cardiovascular fitness. 

\section{Conclusion}

Our modelling study has demonstrated how different aspects of the training programs influence bone adaptation and damage accumulation. We showed how all aspects of a training program contribute to damage, but changes in training duration, training workload, rest duration and rest frequency affected maximum bone damage. It is unlikely that there is one optimal training program that reduces injury risk whilst supporting sufficient racing frequency. Training regimens are dictated by how individual horses adapt to the race speeds and distances they will be required to compete under, therefore the optimum program (or range of programs) may differ between individuals. Mathematical modelling, hence enhances our understanding of biological processes behind bone damage and remodelling, and is a more ethical means of applying this knowledge without interventions or changes to training programs that may be detrimental.

\section*{Data accessibility}
The code associated with this paper is available at https://doi.org/10.5281/zenodo.19173866
\section*{Funding}
This study was funded by the Hong Kong Jockey Club Equine Welfare Research Foundation (MRG-2023-231012) and was in part conducted under the Equine Limb Injury Prevention Research Program funded by Racing Victoria Ltd (RVL), the Victorian Racing Industry Fund (VRIF) of the Victorian State Government, and The University of Melbourne.

\section*{Conflict of Interest}
The authors declare no conflicts of interest.

\bibliography{bibliography}

\appendix
\section{Mathematical model equations}
\label{sec:mathematical_model}

\subsection{Bone adaptation}
The bone volume fraction $f_{BM}$ [dimensionless] evolves over time through the differential equation
\begin{align}
    \frac{df_{BM}}{dt}&= (A_\text{OBL} - A_\text{OCL}) \alpha S_v , \label{eq:dfBMdt}
\end{align}
where $A_\text{OBL}$ and $A_\text{OCL}$ [mm/day] are the osteoblast and osteoclast activities respectively, $\alpha$ [dimensionless]  is the fraction of surface available for remodelling and $S_v$ is the specific surface area. $S_v$ has been found to decrease with $f_{BM}$ in racehorses, and is therefore defined as a function of $f_{BM}$ \citep{lerebours2015relationship}:
\begin{align}
    S_v&=a\sqrt{1-f_{BM}} \cdot [1-b(1-f_{BM})]. \label{eq:Sv} 
\end{align}
Both $A_\text{OBL}$ and $A_\text{OCL}$ are modulated by the intensity of training, and are defined using the Hill functions
\begin{align}
    A_\text{OBL} &= A_\text{OBL}^\text{min} + \frac{(A_\text{OBL}^\text{max} - A_\text{OBL}^\text{min})\psi_\text{tissue}^{\gamma_B}}{(\delta_B^{\gamma_B} + \psi_\text{tissue}^{\gamma_B})}, \label{eq:A_OBL} \\
        A_\text{OCL} &= A_\text{OCL}^\text{min} + \frac{(A_\text{OCL}^\text{max} - A_\text{OCL}^\text{min})\delta_C^{\gamma_C}}{(\delta_C^{\gamma_C} + \psi_\text{tissue}^{\gamma_C})}, \label{eq:A_OCL}
\end{align}
where $A^\text{min}$ [mm/day], $A^\text{max}$ [mm/day], $\delta$ [MPa] and $\gamma$ [dimensionless] are parameters associated with the Hill function. The strain energy density $\psi_\text{tissue}$ [MPa] is defined as
\begin{align}
    \psi_\text{tissue} &=\frac{1}{2E} \sigma^2, \label{eq:psi}
\end{align}
where $E$ [MPa] is the Young's modulus of bone and $\sigma$ [MPa] is the applied stress. Since bone becomes stiffer with both bone volume fraction and strain rate, we set $E$ to
\begin{align}
    E &= E_0 \dot{\varepsilon}^{\gamma_E} f_{BM}^3, \label{eq:stiffness}
\end{align}
where $E_0$ [MPa] is the base stiffness, $\dot{\varepsilon}$ [$s^{-1}$] is the strain rate and $\gamma_E$ [dimensionless] is the exponent of strain rate. 

\subsection{Bone damage}
Damage is defined as a normalised variable between 0 and 1, where a value of 1 corresponds to bone that has failed. The rate of accumulation of bone damage is
\begin{align}
    \frac{dD^*}{dt} = D_f' - D_r', \label{eq:dDdt}
\end{align}
where $D_f'$ is the accumulation rate and $D_r'$ is the repair rate. The damage accumulation rate is given by the equations
\begin{align}
    D_f' &= v_D(\sigma) v_n, 
    \label{eq:damage_formation} \\
    v_D(\sigma) &= 10^{\frac{\sigma-\sigma_0}{\sigma_1}}\frac{E_\text{nom}}{E},\label{eq:v_D}
\end{align}
where $v_D$ is the damage accumulated per stride (as a function of $\sigma$), $v_n$ [day$^{-1}$] is the number of cycles per day and $E_\text{nom}$ [MPa] is the nominal Young's modulus corresponding to bone loading experiments. The parameters $\sigma_0$ and $\sigma_1$ [MPa]are regression coefficients linking bone fatigue life to applied stress.

The damage repair rate is given by
\begin{align}
    D_r' = [A_\text{OBL} + (F_s - 1) A_\text{OCL}]
    \frac{ \alpha S_v D^*}{f_{BM}}, \label{eq:damage_repair}
\end{align}
where $F_s$ [dimensionless] is a specificity factor describing the rate of microcrack removal relative to the removal rate when osteoclasts remove bone at random (where $F_s = 1$).

\subsection{Effect of training on bone remodelling}
Horses train at a speed $s$ and a distance per day $d$. To map these external factors into parameters of the model, the stress $\sigma$, strain rate $\dot{\varepsilon}$ and strides per day $v_n$ are expressed in terms of $s$ and $d$. The speed is associated with a ground reaction force $F_g$ [N/kg body weight] through the relationship
\begin{align}
    F_{g} = 2.778 + 2.1376s - 0.0535s^2,\label{eq:ground_force}
\end{align}
which is subsequently converted to a stress through the relationship
\begin{align}
    \sigma = \sigma_\text{max} \frac{F_g}{F_\text{max}}, \label{eq:joint_stress}
\end{align}
where $\sigma_\text{max} = 90\ \si{MPa}$ and $F_\text{max} = 24.13\ \text{N/kg body weight}$.

The strain rate is related to the speed through the equation
\begin{align}
    \dot{\varepsilon} = 0.01548s + 0.0004955s^2. \label{eq:strain_rate}
\end{align}
Finally, the number of cycles per day is calculated using a combination of speed and distance per day. At a given speed $s$, the stride frequency is
\begin{align}
    f = 1.7052+0.0305s+0.0004s^2 .
\end{align}
Therefore, the number of cycles per day is
\begin{align}
    v_n = df/s.
\end{align}

\subsection{Combination of different speeds}\label{speeds}
Since horses may train at different speeds at various phases of their preparation, we express the external parameters of our model $\sigma$, $\dot{\varepsilon}$ and $v_n$ as functions as time. At any point in time, racehorses train at $N$ speeds $s_1, s_2, \ldots, s_N$, assumed to be ordered from fastest to slowest. The corresponding distances per day for each speed are $d_1, d_2, \ldots, d_N$.

To map training programs into model inputs, we first calculate mechanical parameters associated with each speed:
\begin{align}
    \sigma_i &= \sigma_\text{max} \frac{2.778 + 2.1376s_i - 0.0535s_i^2}{F_\text{max}}, \\
    v_{n,i} &= d_i(1.7052+0.0305s_i+0.0004s_i^2)/s_i, \\
    \dot{\varepsilon}_i &= 0.01548s_i + 0.0004955s_i^2,
\end{align}
where $\sigma_i$, $v_{n,i}$ and $\dot{\varepsilon}_i$ are the stress, cycles per day and strain rate associated with speed $i$. Thus, the damage associated with speed $i$ is
\begin{align}
    D_{f,i}' = v_D(\sigma_i) v_{n,i}.
\end{align}
The stress and strain rate that osteocytes sense are assumed to be those associated with the highest speed trained at that point in time, hence
\begin{align}
    \sigma &= \max_i \sigma_i = \sigma_1, \\
    \dot{\varepsilon} &= \max_i \dot{\varepsilon}_i = \dot{\varepsilon}_1.
\end{align}
The rate of damage accumulation is the sum of damages accumulated at different speeds, and hence \autoref{eq:damage_formation} is replaced by
\begin{align}
    D_f' = \sum_i^N D_{f,i}'.
\end{align}

\subsection{Model parameters}
Parameters of the model are given in \autoref{tab:parameters}.

\begin{table}
\small
    {\centering
\caption{Parameters for the mathematical model of subchondral bone adaptation in Thoroughbred racehorses \citep{Pan2025Apr}.}
\label{tab:parameters}
    \begin{tabular}{clcccl}
    \toprule 
         \textbf{Parameter}&  \textbf{Description}&  \textbf{Units}&  \textbf{Value}&  \textbf{Reference}\\ \midrule
         $\alpha$&  Fraction of specific surface&  --&  0.19 &  \citet{Whitton2013Jul}\\
         $a$&  Fitting parameter relating $f_{BM}$ to $S_v$&  $\si{mm^{-1}}$&  11.42 &  \citet{Hitchens2018Jun}\\
         $b$&  Fitting parameter relating $f_{BM}$ to $S_v$&  --&  $-0.02$ &  \citet{Hitchens2018Jun}\\
         $E_0$&  Maximum bone stiffness&  MPa&  2500 & \citet{Martig2013Dec}\\
         $\gamma_E$&  Exponent of strain rate in stiffness equation&  --&  0.06 & \citet{Carter1977Oct}\\
        $A_\text{OBL}^\text{min}$& Minimum bone formation rate& $\si{mm/day}$& 0.001094& \citet{Firth2005Apr}\\
        $A_\text{OBL}^\text{max}$& Maximum bone formation rate& $\si{mm/day}$& 0.00603& \citet{Pan2025Apr}\\
        $\delta_B$& Half-saturation stress of bone formation& MPa& 7 & \citet{Les1994Nov}\\
        $\gamma_B$& Sigmoidicity of bone formation& --& 1 & \citet{Hitchens2018Jun}\\
        $A_\text{OCL}^\text{min}$& Minimum bone resorption rate& $\si{mm/day}$& 0.001 & \citet{Boyde2005Apr}\\
        $A_\text{OCL}^\text{max}$& Maximum bone resorption rate& $\si{mm/day}$& 0.00358& \citet{Pan2025Apr}\\
        $\delta_C$& Half-saturation stress of bone resorption& MPa& 1 & \citet{Pan2025Apr}\\
        $\gamma_C$& Sigmoidicity of bone resorption& --& 2 & \citet{Pan2025Apr}\\
        $\sigma_0$& Fitting parameter relating stress to fatigue life& MPa& 139.0& \citet{Pan2025Apr}\\
        $\sigma_1$& Fitting parameter relating stress to fatigue life& MPa& 14.1 & \citet{Martig2013Dec}\\
        $E_\text{nom}$& Reference stiffness for stress-fatigue life equation& MPa& 1714.1 & \citet{Pan2025Apr}\\
        $F_s$& Damage repair specificity factor& --& 5 & \citet{Martin1995May}\\ \bottomrule
    \end{tabular}}
\end{table}

\end{document}